\documentclass[aps,pra,superscriptaddress,twocolumn,amsmath,amsfonts,amssymb,floatfix]{revtex4-1}
\usepackage{enumerate}
\usepackage{mathtools}
\usepackage{amsmath}
\usepackage{graphicx}
\usepackage{epsfig}
\usepackage{sidecap}
\usepackage{hyperref}
\usepackage{color}
\usepackage{subfigure,array}
\usepackage{verbatim}

\usepackage{enumerate}
\usepackage{bm}

\begin{document}

\title{Signature of Supersolidity in a Driven Cubic-Quartic Nonlinear Schr\"odinger Equation}

\author{Argha Debnath}
\affiliation{Department of Physics, School of Engineering and Applied Sciences, Bennett University, Greater Noida, UP-201310, India}
\author{Jammu Tarun}
\affiliation{Department of Physics, School of Engineering and Applied Sciences, Bennett University, Greater Noida, UP-201310, India}
\affiliation{Department of Physics, Universit\"at zu K\"oln, K\"oln, Germany}
\author{Ayan Khan} \thanks{ayan.khan@bennett.edu.in}
\affiliation{Department of Physics, School of Engineering and Applied Sciences, Bennett University, Greater Noida, UP-201310, India}

\begin{abstract}
We present analytical solution, which is periodic in nature, for a driven cubic-quartic nonlinear Schr\"odinger equation (DCQNLSE) is placed in a bi-chromatic optical lattice. 
The solution indicate the creation of density wave. Since, beyond mean-field contribution in quasi one dimensional and one dimensional geometry differs on the even exponents of the nonlinearity thus we extend our analysis towards quadratic-cubic-quartic and quadratic-cubic nonlinearities as well. Later, we study the dynamics of DCQNLSE. Our study indicates the existence of stripe phase along with considerable phase coherence. These findings allow us to comment on the possible emergence of supersolid phase in a condensate.
\end{abstract}

\maketitle

\section{Introduction}
Superfluidity and lattice order are mutually exclusive properties. However, there exists the theoretical proposition for the coexistence of such phases in nature which is popularly described as \textit{supersolid}. Though solid $^{4}He$ is long considered as a prime candidate albeit decades of research unable to provide any unambiguous proof in that direction \cite{massimo}. However, a set of recent experiments in ultra-cold atomic gases have actually exhibited the existence of such a counter-intuitive phase featuring antithetic properties \cite{li,donner,tanzi,bottcher,chomaz}. This apparently contradictory phase of matter could yield deeper insights in understanding the superfluids and superconductors with far-reaching implications in the field of superconducting magnets and sensors, as well as efficient energy transport \cite{yukalov}.
 
It is important to note that, in the last couple of years the ultra-cold atomic gas community has witnessed some remarkable experimental observations. One such example being the formation of liquid droplet in dipolar and binary Bose gas \cite{barbut3,cabrera1}. It asserts that the liquid state arises at high densities from an equilibrium between attractive inter-atomic forces and short-range repulsion. However, these newly emerged droplets in ultra-cold and extremely dilute atomic gases do not explicitly follow the common theoretical perception of liquid as predicted by van der Waals \cite{barbut2}. These are purely quantum mechanical in nature and manifest quantum fluctuations \cite{barbut3,barbut1}. The origin of the attractive force is understood in the purview of standard mean-field theory whereas the repulsive force originates from the beyond mean-field correction \cite{sala1}. The underlying theory relies on the Lee-Huang-Yang's (LHY) correction \cite{lee} to the mean-field Gross-Pitaevskii (GP) equation \cite{gross,pitae}.

Chronologically, the emergence of droplet played the role of precursor in achieving the supersolid state in ultra-cold gases. The breakthrough in the alternative medium of cold-atom was realized when the signature of supersolid properties were noted in spin-orbit coupled BEC \cite{li}. Later, supersolid properties were also observed in dipolar Bose-Einstein condensates of lanthanide atoms \cite{donner}. In these systems, supersolidity emerges directly from the atomic interactions where the competition between short-range two-body scattering and long-range dipolar interaction plays detrimental role \cite{tanzi,bottcher,chomaz}. The experiments were performed for dipolar alkali gases of $^{166}\textrm{Er}$, $^{164}\textrm{Dy}$ and $^{162}\textrm{Dy}$. The supersolid phase was observed in a well-defined parameter range in between the regular BEC and the droplet phase \cite{chomaz}. In a very recent experiment the emergence and decay of supersolid state at finite temperature is also been noted \cite{sohmen}.  

The commonality between these two exotic phases arises from the fact that in both cases the beyond mean-field interaction plays a crucial role in stabilizing the system.
From the mathematical perspective, the problem boils down to a nonlinear equation where both odd and even exponent of nonlinearity plays a pivotal role, 
whence the origin of the even exponent arises from the LHY correction. It is interesting to note that, the even exponent is {\textit{two}} for a strictly one-dimensional model whereas {\textit{four}} in a 
quasi-one-dimensional (Q1D) model. Hence, the governing equation of motion in a one-dimensional system can be noted as quadratic-cubic nonlinear Schr\"odinger equation (QCNLSE) \cite{petrov1} 
and for Q1D system it turns out cubic-quartic nonlinear Schr\"odinger equation (CQNLSE) \cite{debnath1,edmonds}. 
It must be noted here that, the sinusoidal modes in a self trapped QCNLSE has 
recently been discussed \cite{parit} apart from its localized counterpart \cite{astra}. On the contrary, in CQNLSE, it is been shown that nontrivial solutions do not exist for $m=0$ where $m$ is the moduli parameter of the cnoidal solutions \cite{debnath2}. Here we must remember that, the cnoidal solutions lead to sinusoidal modes for $m=0$. However, the localized solutions were demonstrated \cite{debnath1}.

Nevertheless, the recent experimental developments motivate us to search for sinusoidal mode in Q1D setup. In this article, we show that it is possible to obtain trigonometric solution in a CQNLSE
when trapped in bi-chromatic optical lattice (BOL) and subjected to a periodic driving force. The motivation to introduce BOL lies in the fact that it is generated by the superposition of 
two optical lattices (OL) of different wavelengths and intensities. By tuning the power and the wavelength of the constituent laser beams, one can create a pure OL when required
and vice-versa, allowing precise control over the shape of the trap profile \cite{ajay}. The driving forces play the role of stabilization of the system. 
Here, we like to note that the stabilization process in a nonlinear system is a rudimentary subject with profound implications in diverse branches of science and technology. One of the fundamental aspect in this context is the theory of Lyapunov \cite{lyapunov} where the stability of solutions near to a point of equilibrium were mainly focused. In recent times we have encountered proposition of noise-driven stabilization of nonlinear differential equations \cite{apple}. In this context, it must be noted that the externally driven, nonlinear Schrodinger equation (NLSE) has been investigated in the context of a variety of physical processes such as Josephson junction, charge density waves, twin-core optical fibres, plasma driven by rf field \cite{raju}. 

Here, the obtained periodic modes leave the signature of the existence of striped phase. Hence, our theoretical model promises a much simpler description to obtain a striped phase with potential supersolid properties. We extend our search of analytical solution for driven quadratic-cubic-quartic NLSE (DQCQNLSE) and driven QCNLSE (DQCNLSE) as well. This allows us to comment on the analytical continuation of the nonlinear system from Q1D to 1D transition. In the later part, we concentrate on the coherent control of the DCQNLSE and investigate the spatio-temporal behavior of supercurrent as well as variation of energy density. 

In precise, we systematically study the static analytical solutions of driven CQNLSE (DCQNLSE), DQCQNLSE and DQCNLSE in an optical lattice landscape in Sec.\ref{model} and comment on the analytical continuation from the Q1D to 1D transition. In Sec.\ref{trap} we discuss the scheme to tackle an additional harmonic confinement and associated coherent control using our analytical scheme. We draw our conclusion in Sec.\ref{con}.

\section{Static Solution}\label{model}
Off late, several investigations were dedicated towards a purely one dimensional (1D) system \cite{petrov1,astra} along with Q1D studies \cite{debnath1,edmonds}. It is worth noting that 1D Bose gas does not support the formation of a condensate and therefore a quasi 1D geometry is widely used where the Bose gas is allowed to expand in an optical waveguide while enabling us to observe exotic structures like the bright soliton trains \cite{salomon,strecker}. Albeit, it is undeniable fact that the Bogoliubov theory correctly predicts the energy of a weakly interacting Bose gas by assuming the existence of condensate in one dimension \cite{lieb1,popov,petrov1}. Hence, we like to recall the prescription of dimensional reduction from 3D to Q1D in brief, before moving further ahead. 
If we define a modified GP equation (including beyond mean-field contribution) in 3+1 dimension as \cite{cabrera1},
\begin{eqnarray}
i\hbar\frac{\partial\Psi}{\partial t}=\left[\left(-\frac{\hbar^2}{2m}\nabla^2+V_{trap}\right)+U|\Psi|^2+U'|\Psi|^3\right]\Psi,\nonumber\\
&&\label{3dbgp}
\end{eqnarray}
where $U$, $U'$ and $V_{trap}$ are the two-body interaction strength, beyond mean-field interaction and the 3D harmonic confinement respectively. One can carry out dimensional reduction in Eq.(\ref{3dbgp}) by employing an ansatz as \cite{debnath1}, 
\begin{eqnarray}
\Psi(\mathbf{r}, t) = \frac{1}{\sqrt{2\pi a_B}a_{\perp}}\psi\left(\frac{x}{a_{\perp}},\omega_{\perp}t\right)\exp\left[{\left(-i\omega_{\perp}t-\frac{y^2+z^2}{2a_{\perp}^2}\right)}\right],\nonumber\\\label{ansatz1}
\end{eqnarray}
where, $a_{\perp}=\sqrt{\frac{\hbar}{m\omega_{\perp}}}$ and $\omega_{\perp}$, $a_B$ are transverse trap frequency and Bohr radius respectively.  The resulting dynamical equation in Q1D
can now be expressed as 
\begin{eqnarray}
i\frac{\partial\psi(x,t)}{\partial t} =   \left[ - \frac{1}{2}\frac{\partial^2}{\partial x^2} + \frac{1}{2} \omega_x^2 x^{2}+\mathcal{G}_1 |\psi(x,t)|^2+\mathcal{G}_2 |\psi(x,t)|^3 \right]\psi(x,t).\nonumber\\\label{bgp}
\end{eqnarray}
Here, $\mathcal{G}_1$ and $\mathcal{G}_2$ describes the mean-field and beyond mean-field interaction strengths respectively, whereas $\omega_x$ defines the longitudinal trap frequency.
Experimentally the transverse trapping frequency is typically set more than $10$ times the longitudinal frequency \cite{salomon}. 
In this process we also assume that the interaction energy of atoms are relatively weak compared to that of kinetic energy in the transverse direction \cite{atre}.

At this juncture, it is also crucial to elaborate on the subtle difference between 1D and Q1D systems and dimensional crossover. In the crossover regime, a Q1D system assumes that $\sqrt{na^3}<1$ where as for a 1D Bose gas, $\frac{1}{|\sqrt{n_{1D}a_{1D}}|}<1$ \cite{petrov2}. $n$ and $a$ stand for the particle density and $s$-wave scattering length whereas $n_{1D}$ and $a_{1D}$ are the density and scattering length respectively in one dimension. The 1D counterpart of the density and scattering length can be noted as, $n_{1D}=nL^2$ and $a_{1D}=-L^2/2\pi a$ where $L$ is the box length in which the system is confined. So, the dimensional crossover can be characterized by a parameter $\eta=naL^2$ \cite{petrov2}. $\eta\sim 1$ can be noted as the crossover and $\eta>>1$ signifies strongly interacting 3D Bose gas while $\eta<<1$ defines a strongly interacting 1D Bose gas \cite{petrov1}. 

Our primary focus in this section is to determine the analytical solutions for time independent nonlinear Schr\"odinger equation (NLSE) with beyond mean-field contribution in Q1D and 1D systems.
This implies that we will hover around $\eta\geq1$ (the exponent of nonlinearity from beyond mean-field contribution is 4) to $\eta\leq1$ where the extended GP equation is a QCNLSE. We assume that when $\eta\sim1$ the contribution of quadratic as well as quartic nonlinearity is reflected in the extended GP equation leading to QCQNLSE. Hence, we start our investigation from a DCQNLSE and then extend the analysis to DQCQNLSE and DQCNLSE (assuming the longitudinal trap frequency is very weak such that $\omega_x\rightarrow0$). However, in Sec.\ref{trap} we assume $\omega_x\neq0$. The systematic analysis can shed some light on the transition from Q1D to 1D system. 

\subsubsection*{Cubic-Quartic NLSE}
We start from a CQNLSE in a bi-chromatic lattice. It is already noted that the presence of two optical lattices of different frequency in same spatial dimension is favourable for the 
formation of supersolid where the effective lattice potential was described as a superlattice \cite{li}. Additionally, we employ a periodic driving force as 
we realize that, to compete with the two-body mean-field interaction we require this contribution. It must be noted that in previous all experimental and theoretical descriptions (except Ref. \cite{parit}), we have seen that the two-body mean-field interaction is pitted against either dipolar or spin-orbit coupled interactions. In our analysis, the external force mimics the alternative force which competes with the regular two body interaction. Nevertheless, we also like to note that, the use of external driving force in ultracold atomic systems is nothing new. There are suggestions for generating
and controlling the transport of BEC atoms from a reservoir to the waveguide via a source/driving force \cite{paul,yan}. Here, the source term actually models the coupling of waveguide with a BEC reservoir. 
Very recently, we observe an analysis of quadratic-cubic NLSE using the source term \cite{cn}.
Apart from these, a wide class of solutions of GP equation in presence of external source has already been studied quite extensively \cite{raju}.

The knowledge of Eq.(\ref{bgp}) allows us to write a generic time-dependent DCQNLSE as,
\begin{eqnarray}\label{cqnlse1}
&&-\frac{1}{2}\frac{\partial^2\Psi}{\partial x^2}+\left(V_2\sin{\zeta x}-V_1\sin^3{\zeta x}\right)\Psi+g_1|\Psi|^2\Psi\nonumber\\
&&+g_2|\Psi|^3\Psi-i\frac{\partial\Psi}{\partial t}=F'(x,t).
\end{eqnarray}
At this moment we focus on the static solution so that $\Psi(x,t)=\psi(x)e^{-i\mu t}$, where $\mu$ is the chemical potential. The driving force is phase locked temporally with the solution and experiences sinusoidal modulation in the spatial dimension such that $F'(x,t)=F e^{-i\mu t}\sin{\zeta x}$. Hence, the time independent DCQNLSE will read,
\begin{eqnarray}\label{cqnlse2}
&&-\frac{1}{2}\frac{d^2\psi}{dx^2}+\left(V_2\sin{\zeta x}-V_1\sin^3{\zeta x}\right)\psi+g_1|\psi|^2\psi\nonumber\\
&&+g_2|\psi|^3\psi-\mu\psi=F\sin{\zeta x}.
\end{eqnarray}
Here, $\zeta$ is the inverse of coherence length, $g_1$ and $g_2$ are the strength of the cubic and quartic nonlinearities respectively. $V_1$ and $V_2$ are potential depths of the periodic traps whose superimposition in the same spacial dimension creates a bi-chromatic landscape which can even be modulated to create a superlattice \cite{li}. 
Frequencies of the two laser beams responsible in creating the BOL are commensurate. We can also express the BOL as, $V_{BOL}=\frac{V_1}{4}(3\sin(\zeta x)-\sin(3\zeta x))-V_2\sin(\zeta x)=(\frac{3V_1}{4}+V_2)\sin(\zeta x)-\frac{V_1}{4}\sin(3\zeta x)$. 
Hence, it suggests that, apart from different amplitude, the wavenumber of one laser is required to be thrice of the second laser. In practice, the superlattice potential was created using two different laser beams associated with two different wavenumber albeit equal amplitude \cite{li}. In a recent numerical study, the atoms were subjected to spin dependent periodic potential \cite{hanwei}.

Further, $F$ is the strength of the periodic driving force. We assume an ansatz solution of the form $\psi(x)=A+B\sin{\zeta x}$ and apply in Eq.(\ref{cqnlse2}). The rationale for choosing this specific type of ansatz is derived from the fact that the experimental observation of the density distribution of the supersolid phase is well fitted through a function of similar form \cite{tanzi}. Using this ansatz, we yield a set of consistency conditions.
\begin{eqnarray}
&&A^2g_1+A^3g_2-\mu=0\label{ceq1}\\
&&6A^2Bg_1+8A^3Bg_2+2AV_2+B\zeta^2-2B\mu-2F=0\nonumber\\\label{ceq2}\\
&&3ABg_1+6A^2Bg_2+V_2=0\label{ceq3}\\
&&B^3g_1+4AB^3g_2-AV_1=0\label{ceq4}\\
&&B^3g_2-V_1=0\label{ceq5}
\end{eqnarray}
\begin{figure}
\includegraphics[scale=0.25]{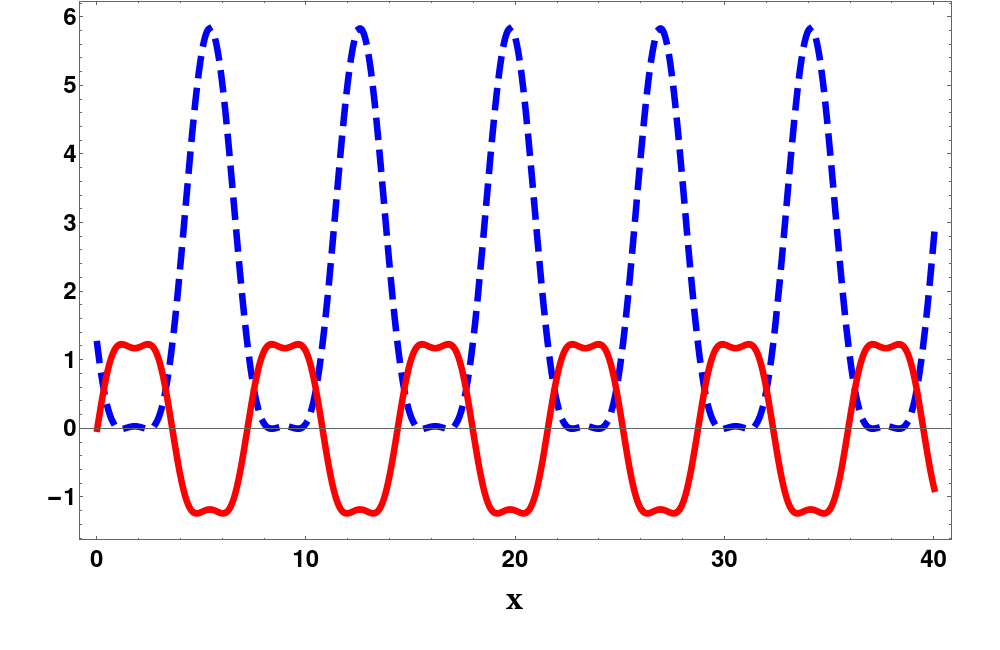}
\caption{(Color Online) The density profile calculated from Eq.(\ref{sol1}) is depicted here along with the bi-chromatic lattice potential. The blue dashed line describes the density variation ($|\psi(x)|^2$), where as the red solid line demonstrates the potential landscape (the bi-chromatic lattice or $V_{BOL}$) in the same spatial dimension. We have used arbitrary units for this plot such that, $V_1=1.0$, $V_2=\frac{1}{3}\left(\frac{V_1}{g_2}\right)^{1/3}\frac{g_1^2}{g_2}$, $g_1=1.5$, $g_2=0.3g_1$, $F=0.5$.}\label{sl}
\end{figure}

A careful analysis of the consistency conditions allows us to yield the exact analytical solution. First, we use Eq.(\ref{ceq5}) to determine $B$ as a function of equation parameters, such as $V_1$ and $g_2$. This results $B=(\frac{V_1}{g_2})^{1/3}$. Using the fact $V_1=g_2B^3$, in Eq.(\ref{ceq4}) we immediately obtain $A=-g_1/3g_2$. Further, from Eq.(\ref{ceq3}) one can determine $V_2$ as a function of interaction strengths which yields $V_2=\frac{Bg_1^2}{3g_2}$. All these information now allow us to explicate the chemical potential ($\mu$) as, $2g_1^3/27g_2^2$ deriving from Eq.(\ref{ceq1}). Lastly we obtain an expression for inverse coherence length or $\zeta$ as, $\zeta=\pm\left(\frac{2F}{B}\right)^{1/2}$ by using Eq.(\ref{ceq2}). 

In Eq.(\ref{ceq1}-\ref{ceq5}), $g_1$, $g_2$, $V_1$ and $F$ are equation parameters whereas $A$, $B$, $\mu$, $V_2$ and $\zeta$ are the solution parameter. In actual experimental setup, $V_2$ can also be treated as known parameter so as $\zeta$. However, we find here that it is not possible to treat $V_2$ and $\zeta$ as independent parameter rather they are coupled to the interaction strengths as well as driving force and $V_1$. Hence, from the experimental perspective, to obtain sinusoidal modes as described, we need to control $V_2$, the amplitude of the second laser, via precise control of the external magnetic field (through Feshbach resonance) and amplitude of the first laser. Similarly, the coherence length is now coupled with the amplitude of the driving force as well as $V_1$ and $g_2$. Though our approach leads to these additional constrain condition however, it must not disregard the fact that the current formalism do suggest sinusoidal modes, which can lead to supersolid like phase, without taking into account dipolar or spin-orbit interactions.  

Finally, we note that solution for Eq.(\ref{cqnlse1}) as,
\begin{eqnarray}\label{sol1}
\psi(x)=\psi_{Q1D}(x)=-\frac{g_1}{3g_2}+\left(\frac{V_1}{g_2}\right)^{1/3}\sin{\zeta x}.
\end{eqnarray}
These findings also suggests that $B\in\mathbb{R}$ to avoid the possibility of complex coherence length. In other words this also implies that the beyond mean-field interaction strength ($g_2$) is repulsive in nature along with the fact that $V_1>0$. However, it must be noted that there is no such restriction on the two-body interaction strength ($g_1$). Hence, the mean-field interaction strength can be attractive as well as repulsive.
The driving force ($F$) must have the same direction as the displacement.

In Fig.~\ref{sl} the spatial variation of density ($|\psi(x)|^2$) is depicted via blue dashed line. The red solid line describes the spatial variation of the BOL. From the figure, the existence of density wave is quite evident with density maxima coinciding with the potential minima. 

It is now important to analyze the stability of these modes. It is well accepted that, for sinusoidal excitation in nonlinear Schr\"odinger-type equations,
the stability can be examined through the Vakhitov-Kolokolov (VK) criterion \cite{vakhitov,das2}. According to this criterion, the stability condition can be assessed based on the sign of the slope for the number of atoms per lattice site ($\mathcal{N}$) with respect to the chemical potential. This implies that, if $\partial\mathcal{N}/\partial\mu>0$ then the solution is stable, conversely for negative slope the solution is unstable and when $\partial\mathcal{N}/\partial\mu=0$ then it is marginally stable. Here, we find that $\partial\mathcal{N}/\partial\mu=\mathcal{L}/g_1$ when we calculate the particle number in a unit cell of length $\mathcal{L}$. Hence, for the solution to be stable, $g_1$ must remain positive or repulsive in nature. It must be noted that in the experiments the stabilization mechanism of the periodic modes can be described by the competition of the repulsive short-range interaction with long-range dipolar interaction \cite{tanzi}. Similarly, in our model the repulsive short-range interaction is balanced by the driving force. 

\subsubsection*{Quadratic-Cubic-Quartic NLSE}
\begin{figure}
\includegraphics[scale=0.25]{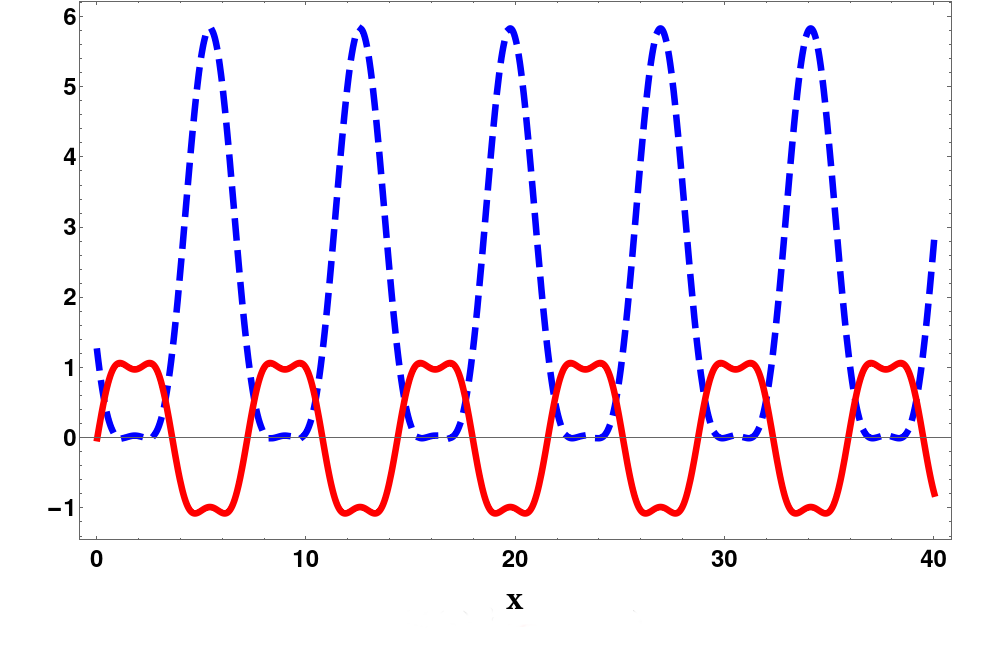}
\caption{(Color Online) The density wave against the backdrop of the bi-chromatic potential landscape for a DQCQNLSE. The blue dashed line describes the density variation ($|\psi(x)|^2$), where as the red solid line demonstrates the potential landscape ($V_{BOL}$) in the same spatial dimension. $F=0.5$, $V_1=1.0$, $g_1=1.5$, $g_2=0.3g_1$ and $g_3=0.1g_1$}\label{qcq_den}
\end{figure}

As promised earlier, we now extend our analysis to a DQCQNLSE with the objective to find a periodic solution. Effectively, we only add a quadratic interaction term in our original Eq.(\ref{cqnlse2}) as it is already agreed upon that in 1D system the beyond mean-field contribution is described by a quadratic nonlinearity \cite{astra}. Hence a forced QCQNLSE or DQCQNLSE can be defined as,
\begin{eqnarray}\label{qcqnlse1}
&&-\frac{1}{2}\frac{d^2\psi}{dx^2}+\left(V_2\sin{\zeta x}-V_1\sin^3{\zeta x}\right)\psi+g_1|\psi|^2\psi\nonumber\\
&&+g_2|\psi|^3\psi+g_3|\psi|\psi-\mu\psi=F\sin{\zeta x}.
\end{eqnarray}
Here, $\zeta$, $F$, $g_1$, $g_2$, $V_1$ and $V_2$ carries their earlier defined meaning. $g_3$ is the strength of the quadratic nonlinearity. Following, the older prescription we assume an ansatz solution of the form $\psi(x)=A+B\sin{\zeta x}$ and apply in Eq.(\ref{qcqnlse1}). The new set of consistency conditions are as follows:
\begin{eqnarray}
&&A^2g_1+A^3g_2+A g_3-\mu=0\label{qc1}\\
&&6A^2Bg_1+8A^3Bg_2+4ABg_3+2AV_2+B\zeta^2-2B\mu-2F=0\nonumber\\\label{qc2}\\
&&3ABg_1+6A^2Bg_2+Bg_3+V_2=0\label{qc3}\\
&&B^3g_1+4AB^3g_2-AV_1=0\label{qc4}\\
&&B^3g_1-V_1=0.\label{qc5}
\end{eqnarray}
It turns out that $A$ and $B$ yields same result as noted above, i.e., $A=-g_1/3g_2$ and $B=(\frac{V_1}{g_2})^{1/3}$. Hence, $\psi(x)=\psi_{Q1D-1D}(x)=-\frac{g_1}{3g_2}+\left(\frac{V_1}{g_2}\right)^{1/3}\sin{\zeta x}$. However, we observe the contribution of $g_3$ in describing the strength of the linear optical lattice potential and the chemical potential such that, $V_2=\frac{B}{3g_2}(g_1^2-g_2g_3)$ and $\mu=\frac{g_1}{27g_2^2}(2g_1^2-9g_2g_3)$. Fig.~\ref{qcq_den} describes the density profile and the spatial variation of the potential. The blue dashed line corresponds to the density wave and the red solid line depicts the potential landscape. The figure is almost exactly similar to Fig.~\ref{sl} as the solution in both the cases remains same.

Next, it is important to comment on the stability of these solutions. Applying the stability criterion as described earlier, we obtain $\partial\mathcal{N}/\partial\mu=\frac{2\mathcal{L}g_1}{2g_1^2-3g_2g_3}$, where $\mathcal{L}$ is the length of the unit cell. So, the VK criterion is applicable iff $2g_1^2\neq3g_2g_3$. Therefore, the solution is stable provided (i) $g_1>0$ and $2g_1^2>3g_2g_3$ or (ii)$g_1<0$ and $2g_1^2<3g_2g_3$. However, for all practical purposes, we expect the first condition to be satisfied as $g_2$ and $g_3$ are expected to be relatively smaller compared to $g_1$. We have already commented that the solution does not allow $g_2$ to be negative albeit there is no such restriction on $g_3$. If $g_3$ is absent then again the stability relation boils down to the same expression as described before.

Very significantly, we observe that the solution does not change in DCQNLSE and DQCQNLSE regimes ($\psi_{Q1D}(x)=\psi_{Q1D-1D}(x)$), thus signifying a smooth transition between these two interaction domains. In precise it implies that, if $\eta\rightarrow1$ from right, the system retains analytical continuation. However, it is also important to note that to achieve this feat, it is required to tune one of the external lattice potential ($V_2$) accordingly. We also observe a subsequent change in the chemical potential from $2g_1^3/27g_2^2$ to $\frac{g_1}{27g_2^2}(2g_1^2-9g_2g_3)$, however we can return to the earlier expression as soon as we assume $g_3=0$. 

\subsubsection*{Quadratic-Cubic NLSE}
On the contrary, to obtain an analytical solution for a driven 1D NLSE we need to switch off the BOL and an OL appears sufficient to support the solution. It is worth noting that in a QCNLSE, it is possible to obtain sinusoidal solution even without any lattice potential or external driving \cite{parit}. However, as a part of our systematic study of transition from Q1D to 1D geometry we intended to employ minimal change in the system. Through our survey, we realize that, a minor trap engineering by means of transforming a BOL to OL is sufficient to obtain the desired result. 

Hence, we consider here the OL as $V_3\sin^2(\zeta x)$. The equation of motion can be noted as,
\begin{eqnarray}
&&-\frac{1}{2}\frac{d^2\psi}{dx^2}-\left(V_3\sin^2{\zeta x}+g_3|\psi|+g_1|\psi|^2-\mu\right)\psi=F\sin{\zeta x}.\nonumber\\\label{qcnlse}
\end{eqnarray}
\begin{figure}
\includegraphics[scale=0.25]{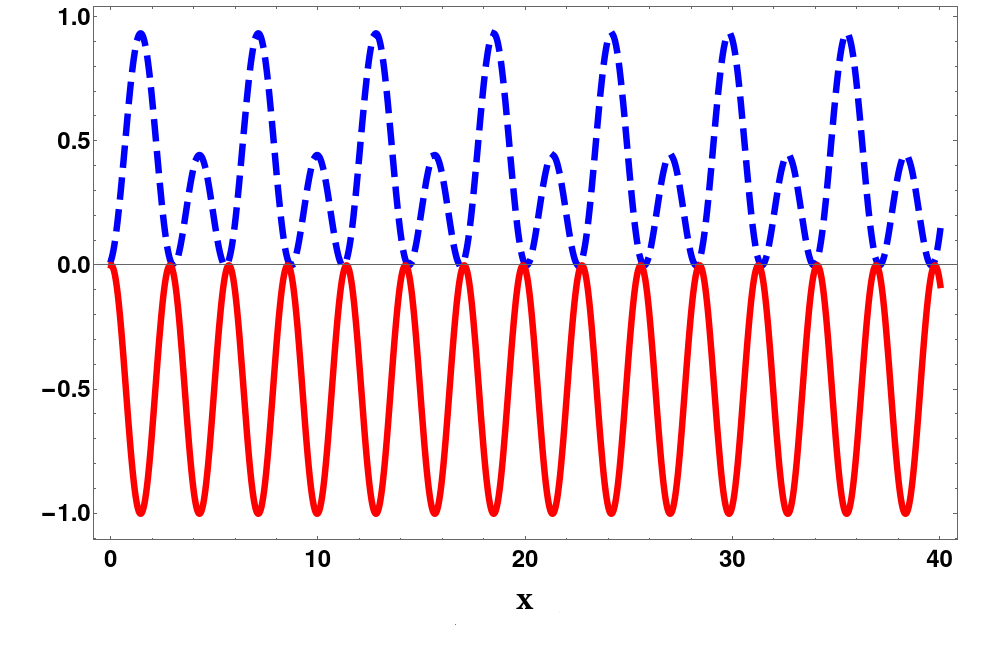}
\caption{(Color Online) The density wave against the backdrop of the optical lattice potential for a DQCNLSE. The blue dashed line describes the density variation ($|\psi(x)|^2$), whereas the red solid line demonstrates the optical lattice potential in the same spatial dimension. $F=0.5$, $V_3=1.0$, $g_1=1.5$ and $g_3=0.1g_1$}\label{qc_den}
\end{figure}

Following the same prescription as earlier in Eq.(\ref{qcnlse}), we determine the wavefunction as $\psi(x)=\psi_{1D}(x)=-\frac{g_3}{2g_1}\pm\sqrt{\frac{V_3}{g_1}}\sin(\zeta x)$ and the chemical potential being $-\frac{g_3^2}{4g_1}$. The inverse coherence length remains same as previous such that, $\zeta=\pm\sqrt{\frac{2F}{B}}$. The density wave is depicted in Fig.~\ref{qc_den} where the blue dashed line described the spatial density variation. The red solid line denoted the optical lattice potential. We skip the detail derivation for brevity, however, we like to note that $\psi_{1D}\neq\psi_{Q1D}$ and thus we lose the analytic continuation as observed till the previous section. 

The V-K criterion of stability of the solution is $\frac{V_3\sin{\zeta\mathcal{L}}}{g_3^2\zeta}-\frac{\mathcal{L}V_3}{g_3^2}>\frac{\mathcal{L}}{g_1}$. It is interesting to note that if $\zeta\mathcal{L}<<1$ then the solution will be stable if $\mathcal{L}/g_1<0$, or the two-body interaction requires to be attractive, which is opposite to the situation described in the previous sections. One may argue that in DQCQNLSE if we insert $g_2=0$ then again we have a DQCNLSE with stability criterion dictating the two-body short-range interaction to be repulsive. However, it must be noted 
$\psi_{Q1D-1D}(x)\rightarrow\infty$ for $g_2=0$ in DQCQNLSE as $B=(V_1/g_2)^{1/3}$ and therefore $g_2=0$ is not admissible. Hence, it is clear that by tuning $\eta$ from right to left will not lead to any smooth transition of the density waves from Q1D to purely 1D. 

From the study of the static solutions, it is evident that periodic density wave do exist, however to comment on the existence of supersolid phase it is necessary to study the dynamics of the solution and subsequent phase coherence. In the next section, we will investigate the analytical solution for a DCQNLSE loaded in a harmonic trap which is tightly confined in the transverse direction and the system is allowed to spread along the longitudinal axis. In the longitudinal direction, the system additionally experiences a BOL. We consider the dynamical solution to have both amplitude and phase. Through a systematic analysis we elaborate the behaviour of the phase and the amplitude. 
  
\section{Dynamic Solution}\label{trap}
In this section, we are interested to obtain an analytical solution derived from the time-dependent DCQNLSE (TDCQNLSE) and investigate on its temporal behaviour. 
We assume the TDCQNLSE with an additional harmonic confinement which is more amenable from the experimental point of view. Also it is worth noting that the use of harmonic trap along with other trapping potentials is a common practice to simulate realistic configurations of cold atoms experiments \cite{hanwei,ajay}. The common practice is to load the ultracold atomic system in a 3D harmonic trap and then carry out necessary modulation based on the experimental demand as in the case of supersolid observation \cite{chomaz}. In the experiment of supersolid the gas was first loaded in a cigar shaped trap which implies tight transverse confinement allowing the atoms to accommodate themselves only in the longitudinal direction ($\frac{1}{2}m\omega_{\perp}^2(y^2+z^2)>>\frac{1}{2}m\omega_x^2 x^2$). As mentioned earlier, typically $\omega_{\perp}/\omega_x\sim 10$ allows us to achieve the cigar-shaped geometry. Here, we plan to explicate the analytical method to treat the TDCQNLSE in a cigar-shaped trap. We observe that the presence of harmonic trap in the
system necessitates the presence of a chirped phase, which yields an efficient nonlinear compression at a desired parameter regime. In presence of a repulsive (regular) harmonic trap, the stripe phases lead to resonances. When the
frequency of the chirped pulses is in resonance with the frequency of the harmonic trap, a significant increase in kinetic energy is observed, which gives rise to the nonlinear compression of the condensate \cite{das}.

\subsubsection*{Coherent Control}
\begin{figure}
\includegraphics[scale=0.35]{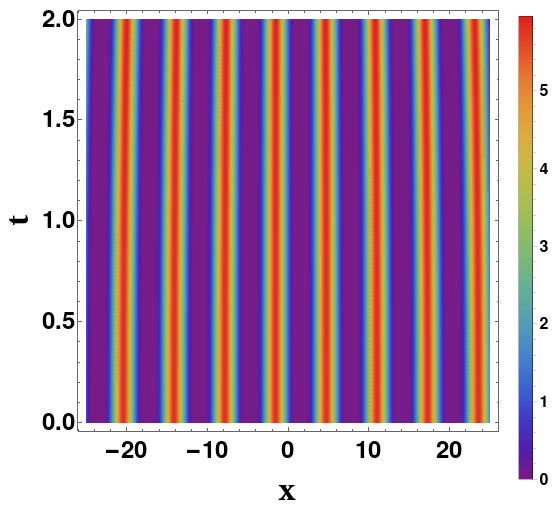}
\caption{(Color Online) Time evolution of the stripe phase in presence of harmonic confinement of angular frequency, $\omega_x=0.1$. The initial amplitude and position of the center of mass is assumed as $1$ and $0.01$ respectively. The potential depth is assumed to be $V_1=1$ and $\Lambda=0.5$. We fix the two-body interaction strength as $g_1=1.5$ and the strength of the beyond mean-field contribution is $g_2=0.3g_1$. The parameter values are arbitrary in nature.}\label{st}
\end{figure}

In our model, with reference to Eq.(\ref{bgp}), we further assume that the longitudinal trap can breathe such that $\omega_x\equiv\omega_x(t)$ and the BOL is super imposed over the pulsating harmonic trap, $\frac{1}{2}\omega_x(t)x^2$, in the same spatial dimension. Additionally, we assume that the interaction strengths can be modulated temporarily such that the two body interaction is defined as $\mathcal{G}_1(t)$
and the beyond mean-field interaction can be defined as $\mathcal{G}_2(t)$. 
Hence, the TDCQNLSE can be described as:
\begin{eqnarray}\label{tcqnlse1}
i\frac{\partial\psi}{\partial t}-\mathcal{H}\psi=\mathcal{F}(x,t),
\end{eqnarray}
where, $\mathcal{H}(x,t)=-\frac{1}{2}\frac{\partial^2}{\partial x^2}+\frac{1}{2}\omega_x^2(t)x^2+\mathcal{V}_{BOL}(x,t)+\mathcal{G}_1(t)|\psi|^2\nonumber+\mathcal{G}_2(t)|\psi|^3-\nu(t)$.
Here, $\nu(t)$ is the temporally modulated chemical potential. We assume an ansatz solution for Eq.(\ref{tcqnlse1}) such that,
\begin{eqnarray}\label{ansatz1}
\psi(x,t)&=&\sqrt{\alpha(t)}\Phi[\xi(x,t)]e^{i[\phi(x,t)+\chi(x,t)]},\nonumber\\
&&
\end{eqnarray}
$\phi(x,t)$ describes the chirped phase and $\chi(x,t)$ is the density dependent complex phase of the solution profile \cite{atre}. We assume that, the chirped phase has a quadratic form 
such that, $\phi(x,t)=a(t)-\frac{1}{2}c(t)x^2$. The ansatz solution is described in the center of mass frame where we incorporate usual Galilean transformation such that $\xi(x,t)=\alpha(t)(x-l(t))$. Applying Eq.(\ref{ansatz1}) in Eq.(\ref{tcqnlse1}) and separating the real and imaginary part of Eq.(\ref{tcqnlse1}) result in two equations, commonly noted as, continuity (imaginary part) and pressure (real part) equation \cite{khan4}. A set of consistency condition emerges from the above mentioned equations such that,
\begin{eqnarray}
&&\frac{d\alpha}{dt}=\alpha c;\,\,cl+\frac{dl}{dt}=\Lambda\alpha;\,\,\mathcal{G}_1(t)=\alpha g_1;\,\,\mathcal{G}_2=\sqrt{\alpha}g_2\nonumber\\
&&\mathcal{F}=\alpha^{5/2}F\sin{\xi(x,t)};\,\,\mathcal{V}_{BOL}(x,t)=\alpha^2V(\xi(x,t));\nonumber\\
&&\nu(t)=\alpha^2\mu
\end{eqnarray}
Here, $\Lambda$ plays a role equivalent to velocity. The presence of the trap necessitates chirping of the phase, which enforces another velocity component. In the absence of the trap, it is appropriate to consider $c(t)=0$ then we can easily derive that $l(t)=\Lambda t$. The equation of motion of the center of mass (COM) can also be noted as
\begin{eqnarray}
\frac{d^2l(t)}{dt^2}+\omega_x^2l(t)=0.
\end{eqnarray}
This implies that the oscillation frequency of the COM is same as the trap. Thus for a regular harmonic oscillator trap with constant $\omega_x$ we obtain $l(t)=l_0\sin{\omega_x t}$.
If the trap is pulsating i.e., $\omega_x\equiv\omega_x(t)$ then the COM follows Mathieu function solution \cite{abra}.

It is possible to determine $c(t)$ from a Riccati type equation \cite{atre} such that $\frac{dc}{dt}-c^2=\omega_x^2$ which can be mapped to Schr\"odinger equation, through a 
transformation $c(t)=-\frac{d\ln{b(t)}}{dt}$. Additionally, it can be shown that $a(t)=\lambda\int_0^t \alpha^2dt'$, where $\lambda$ is an arbitrary constant.
\begin{figure}
\includegraphics[scale=0.35]{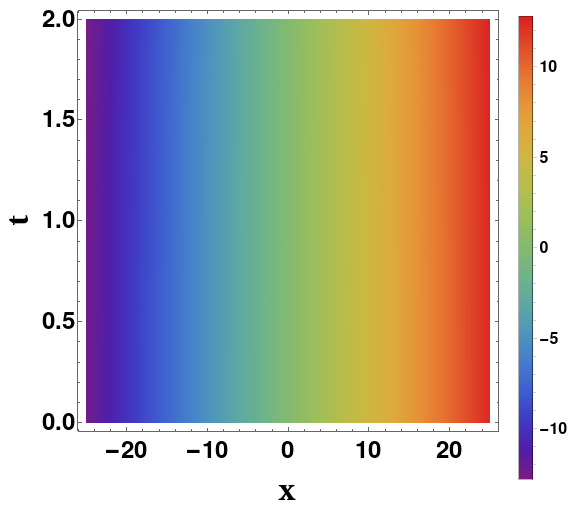}
\caption{(Color Online) The figure depicts the spatio-temporal variation of the phase. The parameter values used for this plot is same as Fig.~\ref{st} and they are arbitrary in nature.}\label{phase}
\end{figure}

The equation of continuity leads to the following phase relation,
\begin{eqnarray}\label{phase_eq}
\frac{d\chi}{d\xi}=\Lambda-\frac{2C_0}{\Phi^2}.
\end{eqnarray}
To avoid the amplitude dependence on the phase, we consider the integration constant $C_0=0$. Hence, we can rewrite the amplitude equation such that,
\begin{eqnarray}
&&-\frac{\lambda}{2}\Phi+\Lambda\Phi\chi_{\xi}+F\sin{\xi}\nonumber\\
&&=-\frac{1}{2}\left(\Phi_{\xi\xi}-\chi_{\xi}^2\Phi\right)+V(\xi)\Phi+g_1|\Phi|^2\Phi+g_2|\Phi|^3\Phi-\mu\Phi.\nonumber\\
\end{eqnarray}
This implies,
\begin{eqnarray}\label{tcqnlse2}
-\frac{1}{2}\Phi_{\xi\xi}+V(\xi)\Phi+g_1|\Phi|^2\Phi+g_2|\Phi|^3\Phi-\tilde{\mu}\Phi=F\sin{\xi}\nonumber\\
\end{eqnarray}
Here, $V(\xi)=V_2\sin\xi-V_1\sin^3\xi$ and $\tilde{\mu}=\left(\mu+\frac{\Lambda}{2}-\frac{\lambda}{2}\right)$. It must be noted here that Eq.(\ref{tcqnlse2}) and Eq.(\ref{cqnlse2}) are of the same structure. Thus an ansatz solution of the form $\Phi(\xi)=A+B\sin\xi$ will yield same result as before. This is a very important aspect of our scheme, which allows the dynamical equation to cast itself in the same form like the static equation. Hence, it is obvious to conclude that the solutions of time-dependent DQCQNLSE and DQCNLSE will have same solution as their static counterpart only differing in the frame of reference of the solutions.
\begin{figure}
\includegraphics[scale=0.35]{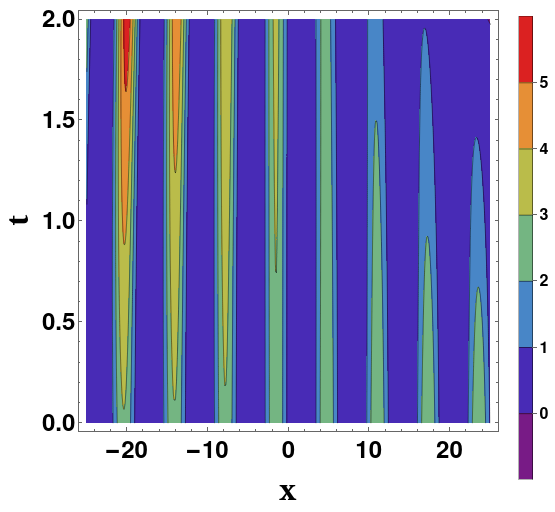}
\caption{(Color Online) The spatio-temporal variation of super current is depicted here. The appearance of supercurrent coincides with the stripes. In the figure the longitudinal trap frequency $\omega_x=0.1$, the lattice depth is considered as $V_1=1$. The amplitude and position of the center of mass at $t=0$ is assumed as $1$ and $0.01$ respectively. The mean-field and beyond mean-field interaction strengths are noted as $g_1=1.5$, $g_2=0.3g_1$. The parameter values are chosen arbitrarily.}\label{current}
\end{figure}

Let us consider the simplest possible situation, where the trap frequency is not pulsating i.e., $\omega_x(t)\equiv\omega_x$. This leads to a trivial solution of the Riccati equation such that $c(t)=\omega_x\tan{\omega_x t}$. Subsequently, one can derive $\alpha(t)=\alpha_0\sec{\omega_x t}$ and $l(t)=\frac{l_0}{\omega_x}\sin{\omega_x t}$. Here $\alpha_0$ and $l_0$ describe the amplitude and position of the COM of the system at $t=0$ respectively. 

The actual solution inside the trap leads to the striped phase as depicted in Fig.~\ref{st}. The corresponding phase can be derived using Eq.(\ref{phase_eq}) and it is been depicted in Fig~\ref{phase}. The moderate spatio-temporal phase variation in the figure points to the phase coherence. The figure clearly suggests that the temporal coherence is retained whereas marginal fluctuations in the spatial coherence. Thus, the existence density wave illustrated via striped phase and possible phase coherence can be attributed to the emergence of supersolidity. Hence, our formalism provide an alternative route to achieve the supersolid phase without using the dipolar BEC, rather applying a calibrated external driving force.

\subsubsection*{Superfluid Current and Density}
It is now quite instructive to calculate the supercurrent as $\mathcal{J}(\xi)=\alpha(t)^2\Phi(\xi)^2\left[-c(t)x+\Lambda\right]$. The calculation of supercurrent plays pivotal role in understanding the dynamical superfluid insulator transition (DSIT). It must be noted here that, DSIT has a classical nature driven by modulational instability and is quite different from the fluctuation-driven quantum phase transition \cite{smerzi,fort1,das1}. 
The behaviour of the supercurrent is shown in Fig~\ref{current}. We observe that the DSIT occurs in this system also, where the atoms transit from
the superfluid phase to an insulating phase, with periodicity analogous to the stripes described in Fig~\ref{st}. The insulating phase wavefunction corresponding to $\mathcal{J}(\xi)=0$ can be noted as $\Phi_I(\xi)=\frac{g_1}{3g_2}\sin{\xi}$. 
\begin{figure}
\includegraphics[scale=0.35]{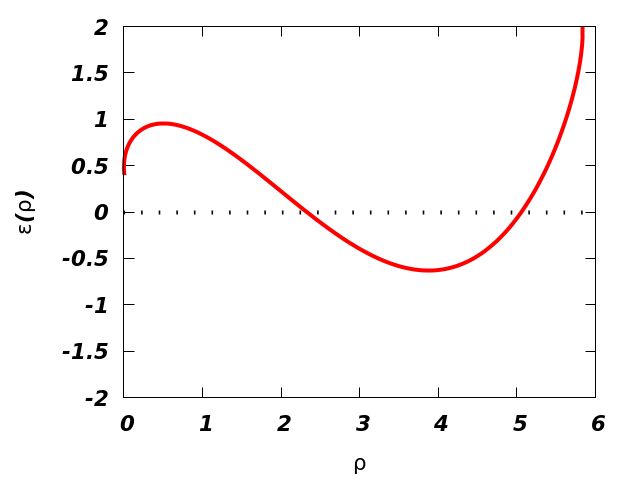}
\caption{(Color Online) The figure describes variation of energy functional as a function of density. The figure is prepared at $t=0$ and we have used $\alpha_0=1$, $l_0=0.01$, $\omega_x=0.1$, $\Lambda=0.5$, $V_1=1$, $g_1=1.5$ and $g_2=0.3g_1$. The values are chosen arbitrarily however they are consistent with the values used to prepare the earlier plots.}\label{ef}
\end{figure}

Off late it has been shown that quantum stabilization results in the formation of quantum droplets and it is possible to form regular arrays of droplets in presence of a trap. However, due to lack of phase coherence they can not be immediately classified as supersolid \cite{tanzi}. Nevertheless the close association of droplet to supersolid is an undeniable fact. In this work we have presented analytical scheme to derive stripe phase solution and explicated their phase coherence. Yet we have not investigated the situation from the prospect of self-bound droplets which additionally show phase coherence. To study the droplet bound state, it is required to calculate the energy functional. The energy functional of the
condensate can be expressed as \cite{brezis}:
\begin{eqnarray}
&&\epsilon(\rho)=\frac{1}{2}\left|\frac{d\psi}{dx}\right|^2+\left(\frac{1}{2}\omega_x^2x^2+\mathcal{V}_{BOL}-\nu(t)\right)\rho\nonumber\\
&&+\frac{1}{2}\mathcal{G}_1\rho^2+\frac{2}{5}\mathcal{G}_2\rho^{5/2},
\end{eqnarray}
where $\rho=|\psi|^2$. 
The terms on the right-hand side, corresponds to the kinetic energy, the
potential energy due to both harmonic and optical lattice confinement. Along with these, a contribution
from the chemical potential interaction potential is also noted. The interaction potential includes usual mean-field contribution as well as the beyond mean-field contribution.
Fig.~\ref{ef} described the variation of energy functional with density at $t=0$. Contrary to the usual energy functional diagram of droplets, we observe that the energy is nonzero at very low density. However, a gradual drop in energy is observed with increase in density. The trajectory clearly suggests the existence of two critical densities of $\rho_{c_1}$ and $\rho_{c_2}$ such that $\rho_{c_1}<\rho_{c_2}$. According to the figure, the energy becomes negative beyond $\rho_{c_1}\sim 2.34$ thereby suggesting formation of self-bound droplets. Further increase in density leads to the equilibrium point where the quantum pressure is zero. This density can be noted as $\rho_{eq}\sim 3.88$ and the energy is lowest at this point. Thereafter the energy starts to grow with increase in density and it becomes positive after $\rho_{c_2}\sim 5.06$ indicating that the scope of bound droplets formation is exhausted beyond this point. Here, one must corroborate with the density favorable for bound state formation with the density described in Fig~\ref{st}. In Fig~\ref{st} we can note that the stripe density varies is varying between $\sim 2$ to $\sim 5.5$ which also quite close to the stripe density range ($\rho_{c_1}\le\rho\le\rho_{c_2}$) where the energy functional is negative as described in Fig~\ref{ef}.
\begin{figure}
\includegraphics[scale=0.35]{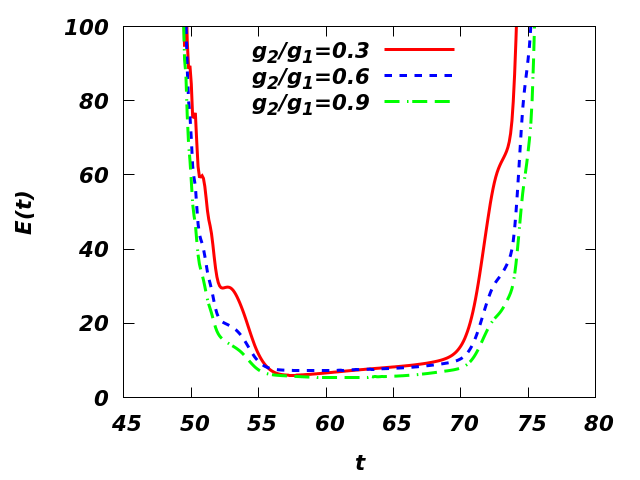}
\caption{(Color Online) The energy spectrum is shown when both harmonic confinement and the BOL is active for different interaction strengths. There is no significant variation due to change in interaction. The parameter values remain same as previous such that $\alpha_0=1$, $l_0=0.01$, $\omega_x=0.1$, $\Lambda=0.5$, $V_1=1$.}\label{eto}
\end{figure}

Now, integrating the energy functional along with the driving force, as noted in Eq.(\ref{Et}), over one period of the lattice potential, we obtain the total energy of the system,
\begin{eqnarray}\label{Et}
E(t)=\int\left(\epsilon(\rho)-\mathcal{F}\right)dx.
\end{eqnarray}
The expression for the energy is
too cumbersome to list and hence, we concentrate on the energy spectrum directly as described in Fig.~\ref{eto}. The condensate shows a rapid
nonlinear resonant increase in energy in the presence of a harmonic and BOL
potential. At certain values of the time variable, it undergoes rapid nonlinear
compression, which in turn mimics the occurrence of resonances in this system. 
These nonlinear resonances occur periodically at the point of nonlinear compression of BEC,
where the density takes its maximum value. The contribution of the quadratic chirped
phase to the kinetic energy is solely responsible for this phenomenon \cite{das2,das}. In other words,
the resonances occur when the driving frequency of the potential and external force matches with the
natural frequency of the system. The resonant behaviour remains same for different interaction ratio between 
$g_1$ and $g_2$. 
This is akin to the observed resonant behaviour in an optical lattice \cite{fabbri}. 
\section{Conclusion}\label{con}
In this article, we have studied DCQNLSE and evaluated its periodic solution which is analytical in nature. The application of this type of system is a Q1D BEC where competition between mean-field and beyond mean-field interaction allows formation of droplets. We then extend our analysis to DQCQNLSE and DQCNLSE to capture the dimensional crossover and effect of beyond mean-field interaction. We observe that a transition from DCQNLSE to DQCQNLSE is smooth where the wave function remains unchanged and the discontinuity is only in the chemical potential. However in DQCNLSE we observe  the wave function itself is different. Thus break down the possibility of any analytical continuation from Q1D to 1D regime. We checked the stability of obtained solutions via VK criterion and noted that the solution remains stable if the two-body mean-field interaction is repulsive. 

Later we explicate the dynamics of DCQNLSE in presence of a harmonic as well as BOL potential. The emergence of stripes becomes quite evident and we study the phase coherence as well. The corresponding energy calculation indicates the existence of critical bound of densities which appears to be  favourable for bound state formation. These critical densities are very similar to the stripe densities. Hence, we can conclude that the stripes point to the supersolid phase. Moreover, we look into the superfluid-insulator dynamical phase transition and study the supercurrent. The spatio-temporal variation of supercurrent closely resembles to the behaviour of the stripes. The energy variation as a function of time demonstrates rapid nonlinear compression, which in turn mimics the occurrence of resonances in this system.

In conclusion, we like to point out that without divulging into condensate with dipolar or spin-orbit interaction we are able to create a scheme, by using an external driving force and a BOL, where one can extract the essence of supersolid phase. We expect our theoretical model will motivate the experimental community to explore the possibility of observing the supersolid phase in a BEC with an additional driving force.  

\section*{Acknowledgement} AK also thanks Department of Science and Technology (DST), India
for the support provided through the project number CRG/2019/000108.


\bibliographystyle{apsrev4-1}
\bibliography{ms_v2}

\end{document}